\documentclass[fleqn,usenatbib]{mnras}
\usepackage{newtxtext,newtxmath}
\usepackage{times}
\usepackage[T1]{fontenc}
\usepackage{ae,aecompl}

\usepackage{graphicx}	% Including figure files
\usepackage{amsmath}	% Advanced maths commands
\usepackage{amssymb}	% Extra maths symbols
\usepackage[flushleft]{threeparttable}
\usepackage{color}
\usepackage{natbib}
\usepackage{array}
\usepackage[normalem]{ulem}

\newcommand\gaia{{\em Gaia}}
\newcommand\solarmass{$M_\odot$}
\newcommand\Rsun{$R_\odot$}
\newcommand\rcore{$r_c$}
\newcommand\mbh{$M_{\rm BH}$}

\newcommand\er{$\pm$}

\newcommand\vnk{$v$\sub{nk}}

\newcommand\porb{$P_{\rm orb}$}
\newcommand\z{$z$}

\newcommand{\sub}[1]{\ensuremath{_{\textrm{#1}}}}
\def\ltsim{\mathrel{\hbox{\rlap{\hbox{\lower3pt\hbox{$\sim$}}}\hbox{$<$}}}}
\def\gtsim{\mathrel{\hbox{\rlap{\hbox{\lower3pt\hbox{$\sim$}}}\hbox{$>$}}}}
\title[On the origin of black hole binaries]{A period-dependent spatial scatter of Galactic Black Hole Transients}
\author[Gandhi et al.]{P. Gandhi,$^{1}$\thanks{E-mail: poshak.gandhi@soton.ac.uk}
A. Rao,$^{1}$
P. A. Charles,$^{1}$ K. Belczynski,$^{2}$ T. J. Maccarone,$^3$ K. Arur,$^3$\newauthor J. M.  Corral-Santana$^4$
\\
$^{1}$Department of Physics \& Astronomy, University of Southampton, Highfield, Southampton SO17 1BJ, UK\\
$^{2}$Nicolaus Copernicus Astronomical Center, Polish Academy of Sciences, ul. Bartycka 18, 00-716 Warsaw, Poland\\
$^{3}$Department of Physics \& Astronomy, Box 41051, Science Building, Texas Tech University, Lubbock, TX 79409-1051, USA\\
$^{4}$European Southern Observatory, Alonso de Cordova 3107, Vitacura, Santiago 19001, Chile
}

\date{Submitted 2020 Feb 06, Revised 2020 Apr 28, Accepted 2020 Apr 29}

\pubyear{2020}

\begin{document}
\label{firstpage}
\pagerange{\pageref{firstpage}--\pageref{lastpage}}
\maketitle

% Abstract of the paper
\begin{abstract}
There remain significant uncertainties in the origin and evolution of black holes in binary systems, in particular regarding their birth sites and the influence of natal kicks. These are long-standing issues, but their debate has been reinvigorated in the era of gravitational wave detections and the improving precision of astrometric measurements. 
Using recent and archival characterisation of 
Galactic black hole X-ray binaries (BHXBs), we report here an apparent anticorrelation between \porb{} (system orbital periods) and scatter in \z\ (elevation above the Galactic plane). The absence of long period sources at high $z$ is not an obvious observational bias, and 
two possible explanatory scenarios are qualitatively explored: (1) a disc origin for BHXBs followed by natal kicks producing the scatter in $z$, with only the tightest binaries preferentially surviving strong kicks; (2) a halo origin, with \porb\ shortening through dynamical interactions in globular clusters (GCs). For the latter case, we show a correspondence in $z$-scatter between BHXBs and the GCs with most compact core radii of $<$\,0.1\,pc. 
However, the known absence of outbursting BHXB transients within Galactic GCs remains puzzling in this case, in contrast to the multitude of known GC neutron star XRBs. 
These results provide an interesting  observational constraint for any black hole binary  evolutionary model to satisfy. 
\end{abstract}

% Select between one and six entries from the list of approved keywords.W

%%%%%%%%%%%%%%%%%%%%%%%%%%%%%%%%%%%%%%%%%%%%%%%%%%

%%%%%%%%%%%%%%%%% BODY OF PAPER %%%%%%%%%%%%%%%%%%

\section{Introduction}
Stellar-mass black holes are the end points of massive star evolution. Observationally, black holes (BHs) have primarily been studied as members of binary systems, either in X-ray binaries (hereafter, BHXBs) or in binaries composed of two compact objects (hereafter, BHBs, referring generically to all BH+[other compact object] binaries). Several distinct channels for formation and growth of these binaries have been proposed, including field formation (e.g. \citealt{belc_apj}), formation in active galactic nuclei discs (e.g. \citealt{bartos17}), in galactic nuclei (e.g. \citealt{oleary09, hamers18}), and dynamical assembly in globular clusters (GCs; e.g. \citealt{rodriguez16}). 
Lack of proper understanding of progenitor nature, supernova explosion physics and natal environment, all hamper efforts to distinguish between these channels.

With respect to environmental uncertainties, GCs have long been known to host neutron star (NS) XRBs, and suggested formation mechanisms include tidal capture of compact objects by ordinary stars \citep{fabian75}; three body encounters \citep{clark75}; direct encounters \citep{verbunt87}; and exchange of companions with primordial binaries either by a black hole or neutron star \citep{hills76}. 
Recently, candidate BHs have also been reported in Galactic GCs,  e.g. M22 \citep{strader12}, M62 \citep{chomiuk}, 47 Tuc \citep{miller15, bah}, NGC 3201 \citep{gie18}. However, none of these have shown outbursts so far, as opposed to many cases of known NSXRB transients in GCs. There are arguably stronger candidates of BH transients in {\em extragalactic} GCs  (e.g., NGC\,4472, \citealt{mac07, maccarone11}; NGC\,1399 \citealt{shih10}; and NGC\,3379,  \citealt{brassington10}), though some of these could be more massive (intermediate mass) black holes (e.g., \citealt{irwin10}).

\vspace*{-0.1cm}
With respect to progenitor physics, the uncertain impact of the `natal kick' velocity (\vnk) imparted to the BH during its formation is of particular importance. Kicks can result either from symmetric mass loss \citep{blaauw61}, neutrino emission asymmetries, or hydrodynamic instabilities in the ejecta during fallback of material onto a short-lived intermediate NS (e.g. \citealt{fryer06,janka17}). 
Natal kicks strongly impact binary survivability and merger rates. For instance, \citet{belc17} demonstrate that for BHBs, a change in the magnitude of \vnk\ by $\sim$\,100\,km\,s$^{-1}$ can change merger rates, and hence gravitational wave (GW) source detection rates, by factors of $\gtsim$\,30. Asymmetric natal kicks should also significantly scramble the locations of BH sources; if strong kicks operate, BHs ought to be efficiently ejected from GCs, but there remains lack of consensus on this issue (e.g. \citealt{strader12, rodriguez16}).

The recent detection of gravitational waves (GW) has reenergised the debate on BH formation channels from observations of extragalactic mergers (e.g. \citealt{belc} and references therein). Contemporaneously, ever larger samples of Galactic BH XRBs are pushing the frontiers of Galactic BH system characterisation \citep{blackcat, watchdog}. Early pioneering works investigated the influence of natal kicks and natal sites by via spatial or kinemetic studies \citep[e.g.][]{vanparadijswhite95,jonker04, repetto12, repetto17, mirabel17-review}, but were naturally limited by small sample statistics. 

Some of these handicaps are being overcome with data from new astrometric surveys which are transforming the understanding of BHs  in Galactic binary systems \citep[e.g. ][]{gaiadr2, gandhi18,atri19, rao19}. Here, we investigate updated samples of Galactic BHXBs to uncover an intriguing relation between the latitudinal height scatter and system orbital periods for BHXBs. We discuss implications of the observed trend with regard to BH formation scenarios. 

\vspace*{-0.5cm}
\section{Sample Analysis and Results}
We began with the complete list of 60 likely BHXB transients presented in the BlackCAT catalogue \citep{blackcat} as of 2019\,October, and compiled their orbital periods (\porb) and distances ($d$) where available. Updates to this sample included: the inclusion of a new dynamically confirmed BHXB MAXI\,J1820+070 at $d$\,$\approx$\,3\,kpc \citep{gandhi18, torres19, atri20}; new likely lower limits of $d$\,$\gtrsim$\,6\,kpc and 5\,kpc for Swift\,J1357.2--0933 and GX\,339--4, respectively \citep{charles19, heida}; and $d$\,$\lesssim$\,25\,kpc for BW\,Cir as a likely upper limit \citep{gandhi18}. It should be stressed that none of these updates substantially impact the sample results presented below. 

\begin{figure*}
	\centering
\begin{minipage}{0.85\textwidth}
%\centering
\hspace*{-1.cm}
\includegraphics[angle=90,width=\linewidth]{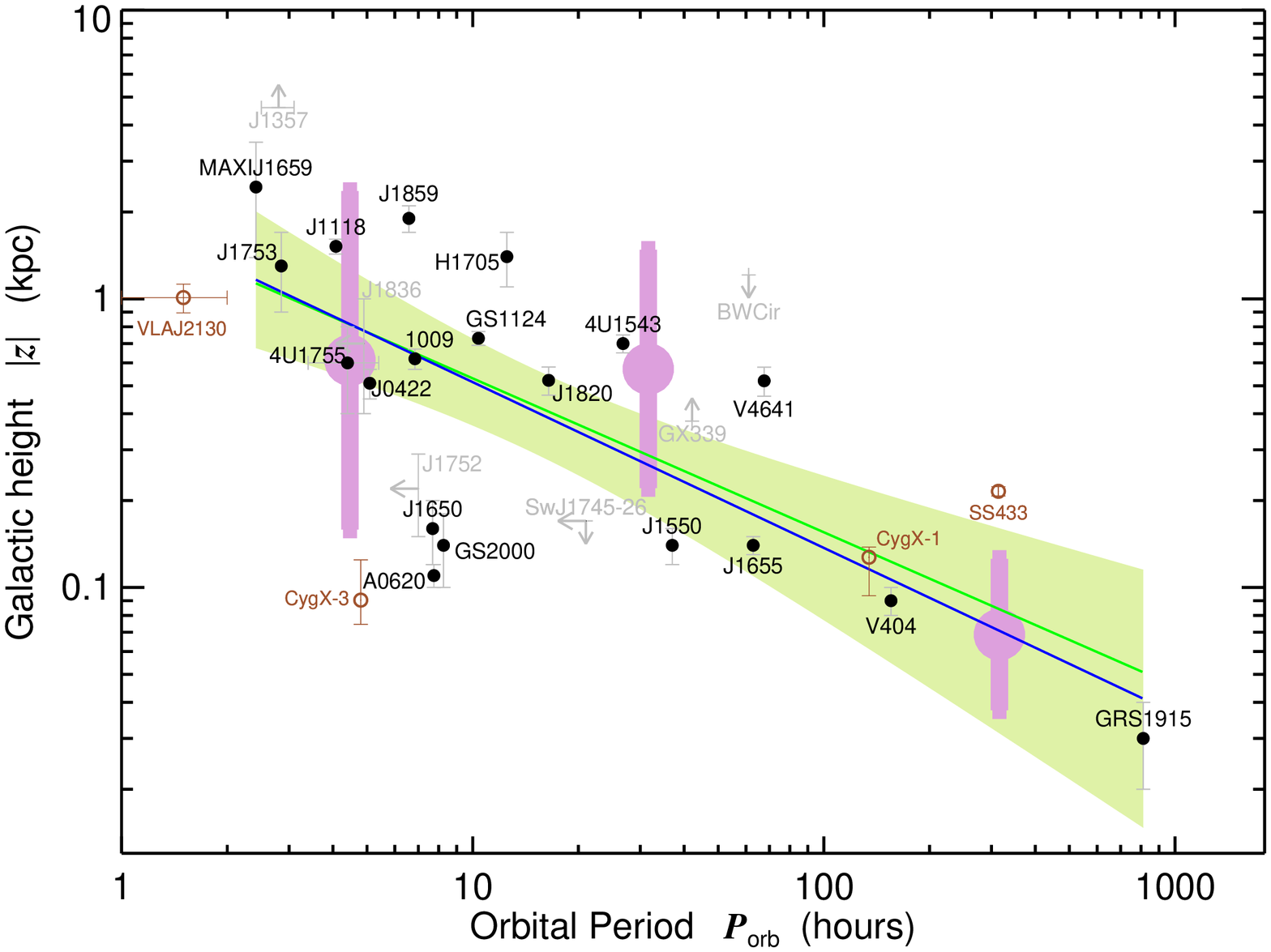}
\end{minipage}
\hspace*{-2.8cm}
\begin{minipage}{0.29\textwidth}
\centering
\includegraphics[width=1.1\linewidth]{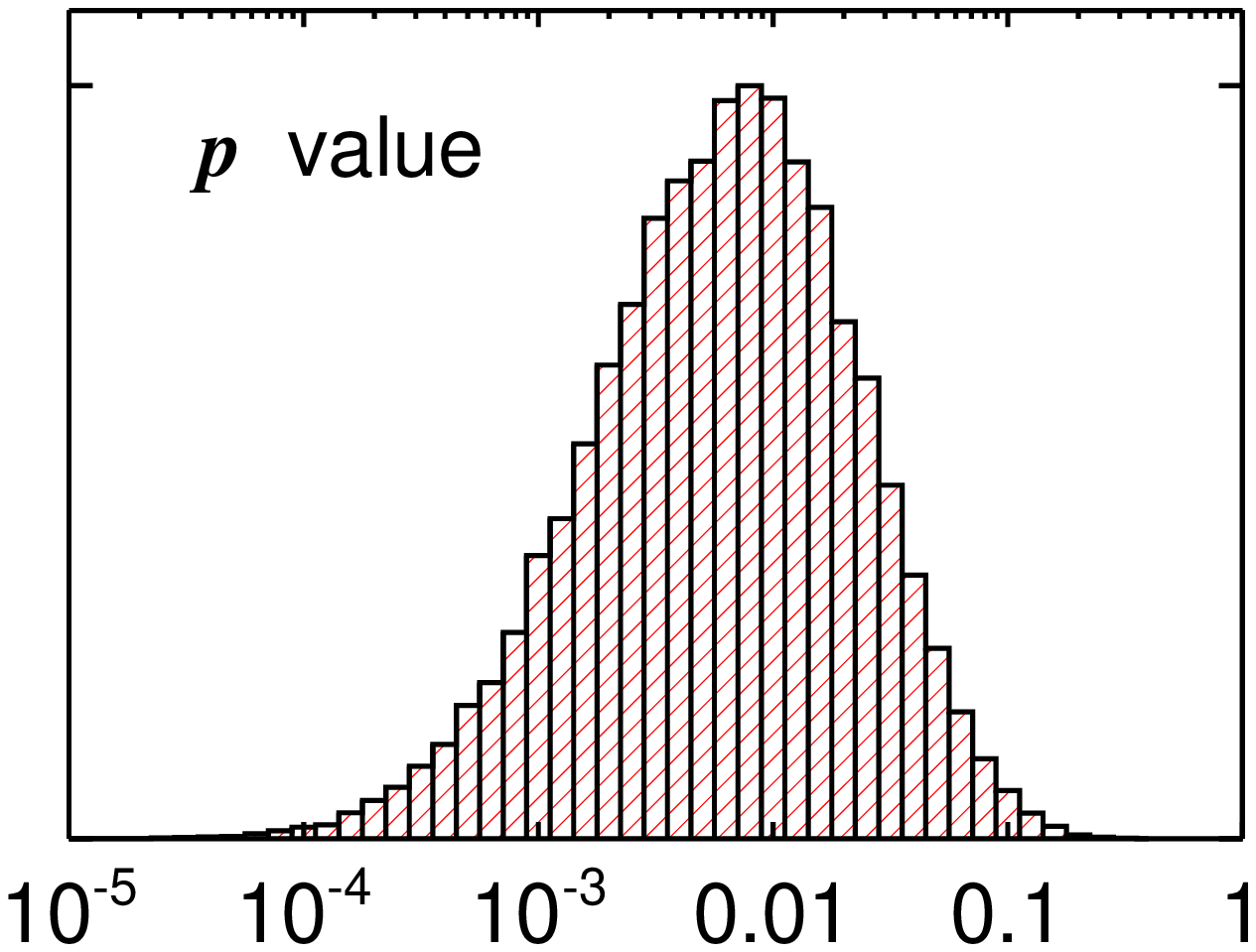}
\includegraphics[width=1.1\linewidth]{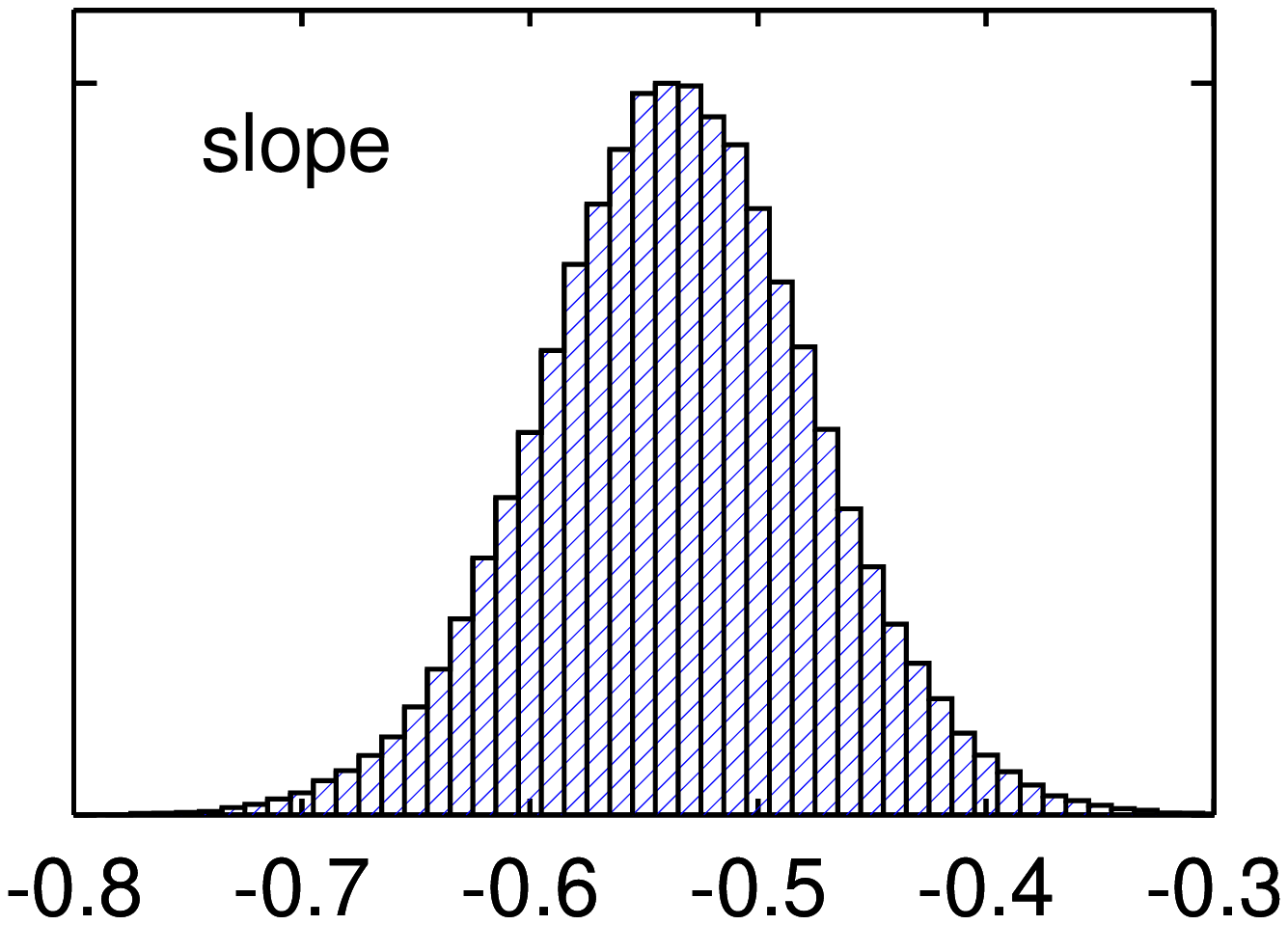}
\end{minipage}
%\vfill
\caption{Height \z{} above the Galactic plane, shown as a function of \porb{} for known BHXBs. Sources with limits 
are plotted in grey. The blue line shows the  regression  
excluding limits, while the green line and hatched region depict the correlation for all objects including limits with 10$^5$ randomised ensembles and the corresponding 95\,\% confidence region, respectively. The thick light purple points with uncertainties depict $z_{\rm rms}$ per dex of \porb. Objects plotted for comparison but not included in the fits are the open brown circles; see text. (Insets on right:) Distributions of null hypothesis probability and  regression slope from the random ensembles.
	\label{porbzname}}
\end{figure*}

Constraints or measurements on the height above the Galactic plane ($|z|$) and the orbital period (\porb) are available for 25 objects. %Of these, 18 are dynamically confirmed systems. 
Fig.\,\ref{porbzname} displays the sample on the log\,$z$--log\,$P_{\rm orb}$ plane, showing that sources are not uniformly distributed in this plane.

% ******************************************************
% ******************************************************

There is a complete absence of long \porb\ systems (\porb\,$\gtrsim$\,1\,day) at high elevations ($\gtrsim$\,1\,kpc), in contrast to the more uniform $z$ filling at shorter periods. 
In other words, the scatter in elevation ($z_{\rm rms}$) decreases progressively with increasing \porb. 
This holds when accounting for limits, and manifests as an apparent anti-correlation between $z$ and \porb. 
A power-law regression  ($z$\,$\propto$\,\porb$^\alpha$) yields a logarithmic slop of $\alpha$\,=\,--0.57\,$\pm$\,0.14 for the 19 BHXBs with measured definite $z$ and \porb. The dispersion around the regression line is entirely dominated by the intrinsic $z$ scatter of 0.37 dex, much larger than the individual statistical uncertainties.

In order to incorporate sources with limits, we created 10$^5$ random ensembles with resampling. Statistical uncertainties on $z$ and \porb, where known, were assumed to represent the standard deviation for a Normal distribution resampling of individual objects. For limits, a uniform distribution was simulated with thresholds based upon reasonable physical assumptions drawing upon the empirical distribution of the detected sources and the size of the Galaxy: $P_{\rm orb}^{\rm min}$\,=\,2\,h, $z_{\rm min}$\,=\,0.01\,kpc, and $d_{\rm max}$\,=\,25\,kpc. None of the results are particularly sensitive to tweaking these assumptions. The median regression slope from 10$^5$ fits to this ensemble is $\alpha$\,=\,--0.54\,$\pm$\,0.06. The ensemble median $p$ value is 0.006$^{+0.017}_{-0.002}$. 
Histograms of these distributions are shown in Fig.\,\ref{porbzname}. 

To summarise the analysis: despite the relatively small current sample size of BHXBs with useful measurements or constraints, there is strongly suggestive evidence of a dependence of $z$ scatter %(or, at least, the scatter in $z$) 
as a function of \porb, based upon several statistical tests. 

The locations of a few other well known sources, that are not part of the transient low-mass BHXB population of BlackCAT, are shown in Fig.\,1 for comparison, but not included in the fit. Two are well known {\em high}-mass XRBs Cyg\,X--1 and SS\,433 which cannot have migrated far from their natal birth sites. The nature of the compact object in both SS\,433 and Cyg\,X--3 remains uncertain. Finally, VLA\,J2130+12 is a new halo BHXB, but its \porb\ remains poorly constrained. These are discussed individually in later sections and cited in the Appendix. But we stress that the locations of all these objects are consistent with the $z_{\rm rms}$ per \porb\ dex (see Fig.\,1), and also consistent with the inference of $z$ scatter decreasing with \porb.

\vspace*{-0.3cm}
\section{Discussion}

To our knowledge, this is the first  report and detailed analysis of a relation between two measureables connecting the {\em spatial} ($z$) and the {\em system} ($P\sub{orb}$) parameters of BHXBs, respectively. Previous works have investigated connections between various spatial parameters (e.g. $z$ and $R_{\rm galactocentric}$; \citealt{jonker04, repetto12}), or additionally against kinematic parameters that can be harder to measure (e.g. natal kick velocities; \citealt{repetto17}). Tentative trends have been noted (e.g. between spectral type and $z$, \citealt{repetto17}; for individual source groups, \citealt{blackcat}; and for \vnk\ and $z$, \citealt{atri19}), but none has been reported as robust. By contrast, both $z$ and $P\sub{orb}$ are relatively easier to measure, requiring a distance estimate and (at minimum) photometric monitoring to measure periodicities. Interpretation could be complex (as discussed below), but the apparent strength of the correlation represents an interesting constraint for any physical model. 

\vspace*{-0.3cm}
\subsection{Potential Biases}
\subsubsection{Observational selection and incompleteness}
We begin with a few words on potential biases. The most significant observational bias is expected to be against short \porb\  systems close to the Galactic plane occupying the lower-left-hand part of Fig.\,\ref{porbzname}, as these are expected to be intrinsically faint and highly extincted. However, there are at least a handful of well characterised systems in this region (e.g. A0620--00, GS\,2000+251, XTE\,J1650-500), and three sources (Swift\,J1745.1--2624, XTE\,J1752--223, MAXI\,J1836--194) whose upper limits could allow them to occupy this regime. In fact, these objects are responsible for the rise in $z$-scatter at short \porb. So while this bias will result in missing sources, it is not critical in terms of explaining the distribution in Fig.\,\ref{porbzname}. 

Instead, the primary cause of the apparent anti-correlation is the lack of objects in the upper-right-hand part of Fig.\,\ref{porbzname}. However, objects in this regime have high $P\sub{orb}$ ($\gtrsim$\,1\,day) and moderate-to-large heights ($z$\,$\gtrsim$\,1\,kpc). Neither effect ought to give rise to an observational bias. Accreting sources with high $P\sub{orb}$ will host large accretion discs, which ought to be bright and easily detectable when and if they undergo outburst, and high $z$ sources will suffer less extinction. So the sample is not obviously biased in a way that could spuriously cause the trend seen in Fig.\,\ref{porbzname}. 

The base sample itself is the BlackCAT catalogue, which is comprehensive but in no way complete in terms of any physical parameter. We have considered the data on all Galactic BHXB candidates so far with constraints on \z{} and $P$\sub{orb}. This probably biases us towards sources that are more amenable to optical follow-up and monitoring; in other words, optically luminous and less extincted systems ought to be well sampled. But again, such systems are preferentially expected to lie in the regime 
where Fig.\,\ref{porbzname} shows an apparent dearth of objects.

Finally, we also checked for the influence of distance, by investigating $P\sub{orb}$ as a function of heliocentric distance. No statistically significant trend was found. 

\subsubsection{The intrinsic \porb\ distribution}
Progenitor systems of accreting binaries are likely to span a 
wide range of orbital periods ranging over orders of magnitude larger than the $P\sub{orb}$ range considered here \citep[e.g. ][]{han98}. However, a variety of effects including mass loss, magnetic braking and gravitational radiation are expected to significantly modify the \porb\ distribution as sources evolve through the accreting binary regime, and the intrinsic distribution of these systems in terms of \porb\ still remains uncertain \citep{arur18}. So it is relevant to ask whether our results are simply a reflection of varying intrinsic population sizes as a function of \porb, exacerbated by the current small samples.% number statistics. 

We tested this in two ways: (1) based upon the {\em model} distributions and expected detectability of transients, and (2) {\em empirically}. For the theoretical test, we began with the code of \citet{arur18} to simulate a range of outburst timescales combined with X-ray detection fractions and optical characterisation likelihood. The \porb\ distribution is 
sampled uniformly between 0--20\,h and independent of $z$. The Galactic BHXB spatial distribution is weighted according to stellar densities, and a three dimensional model incorporating dust reddening as well as gas obscuration was incorporated (see Ibid. for details). 10$^5$ systems were simulated and their detection fraction was computed integrated over small increments of $z$. The result is that BHXB detection probability increases monotonically with \porb, peaking at a healthy $\approx$\,40\,\% for \porb\,=\,20\,h.  %$z$\,$\approx$\,1.5\,kpc.  
This is as expected with large discs being more amenable to detection and bright donor stars being more amenable to follow-up, implying a deficit in the fraction of known long \porb\ systems relative to expectation. 

However, \porb\ values longer than 20\,h were not simulated in the code of \citet{arur18} because their study focused on periods below a well known evolutionary donor-dependent bifurcation in the \porb\ distribution  \citep[e.g. ][]{podsi}. Intermediate mass donor systems are expected to evolve towards longer \porb\ on fast timescales of $\sim$10$^{5-7}$\,years before becoming detached. Thus the relative numbers of systems in this regime could be low, even if their detection is not hampered due to observational biases. 

Therefore, we carried out a second {\em empirical} test to  evaluate how strongly our results are affected by small number statistics at the longest periods. 
This test is essentially model-free, with the only assumptions being that the intrinsic $z$ distribution is well sampled by the current data, and that $z$ is independent of \porb.

One can then compute the likelihood of finding two objects (V404\,Cyg, GRS\,1915+105; the only two at \porb\,$>$\,100\,h), {\em both} in the very low $z$ regime. This is a simple exercise of making two random draws from a distribution (we simulated 10$^5$ ensembles as before, including limits, for which $\bar{|z|}$\,=\,0.54\,kpc with a dispersion of 0.53\,dex). The joint probability of drawing V404\,Cyg and GRS\,1915+105 at their observed low $z$ is found to be 0.002$_{-0.001}^{+0.002}$, with the uncertainties denoting the standard deviation over the ensembles. In other words, the absence of high $z$ long period systems is currently significant at better than 99.6\,\%. Inclusion of Cyg\,X--1 and SS\,433 within the statistics does not change the final inference. Enhancing the census of such systems should be a priority for future surveys, as their expected low extinction could provide excellent opportunities for follow-up.

\subsection{Scenario 1: Disc origin with strong natal kicks}

If BHXBs originate within the Galactic disc, they must evolve to match the observed $z$ distribution today. 
Natal kicks could provide the requisite mechanism for this, although models do not strongly constrain the magnitudes of these  kicks because of unknown dependences on progenitor properties and the complex details of explosion physics. 
As opposed to the LIGO/Virgo population of BHBs, XRBs host only a single compact object, and thus their motion and evolution is impacted by a {\em single} kick. It is the first kick that primarily governs binary survivability and hence merger rates (because binaries are likely less tightly bound at earlier first kick evolutionary stages; e.g. \citealt{belc17}), so understanding natal kicks in BHXBs can inform mass dependent scalings to the BHB regime. 

Observationally, it has been shown that asymmetric progenitor explosion kicks (broadly ranging over 100--500\,km\,s$^{-1}$), in addition to Blaauw kicks, are required for at least some systems in order to account for the present elevations of low-mass BHXB transients above the Galactic plane. This was based upon work done by \citet{repetto12} using Galactic population simulations, accounting for a variety of kick distributions and the location of objects within the Galactic potential. More recent proxies of \vnk\ for BHXBs fall around $\sim$\,80--130\,km\,s$^{-1}$ \citep{gandhi18, atri19}, and this is also the case for the distribution inferred from the first set of more massive BHBs studied with LIGO/Virgo \citep{wong19}. However, accounting for realistic initial spatial and kinematic scatter of progenitors in the disc can greatly mitigate the necessity of strong natal kicks in many cases \citep{belc_apj}, thus complicating proper evaluation of this scenario. Larger samples of XRBs will need to be found and identified before this issue can be settled. Upcoming massive sky surveys should help in this regard \citep[e.g. ][]{johnson19}. 

Once a natal kick has been imparted, its influence on the resultant systemic peculiar motion (as opposed to a change in other degrees of freedom, such as orbital angular momentum) also needs to be understood. In order to gain an appreciation of this, we used the binary population synthesis code {\tt StarTrack} \citep{startrack}, evolving a population of 10$^5$ binaries with system characteristics relevant to BHXB transients (\mbh\,=\,8\,\solarmass, $M_{\rm donor}$\,=\,1\,\solarmass, in a close circular orbit with separation $a$\,=\,20\,\Rsun). Starting with a Maxwellian distribution for the natal kick ($\sigma_{\rm 1D}$\,=\,130\,km\,s$^{-1}$ ), we find that 70\,\%\ of binaries remain bound following BH formation. Their mean \vnk\,=\,180\,km\,s$^{-1}$ (in 3D). Most ($\approx$\,90\,\%) of the kick is transferred to the systemic binary motion ($v_{\rm sys}$\,$\approx$\,161\,km\,s$^{-1}$). Though this is only a single test, it suggests that natal kicks can efficiently translate into high peculiar motion (and thus into a larger $z$ scatter). Full population synthesis tests in the future can use our observed $z$--\porb\ trend together with observed \vnk\ distributions to fully constrain binary survivability.

Finally, we note the location of Cyg\,X--1, a high mass BHXB born in and unambiguously belonging to the thin disc, that likely suffered only a mild natal kick \citep{rao19}. Its position in Fig.\,\ref{porbzname} is entirely consistent with the trend for low-mass transients. The same inference holds for SS\,433, which is associated with the W50 supernova remnant nebula, and could not have meandered substantially from its natal site \citep[e.g. ][]{margon84}.

\subsection{Scenario 2: Globular Cluster Origin}

If GCs are the birth sites of BHXBs, we may expect a correspondence between the spatial scatter of both populations. There are several sources located at high Galactic elevations |$z$|\,$\gtsim$\,1\,kpc (XTE\,J1118+480, H1705--250, XTE\,1859+226, MAXI\,J1659--152, Swift\,J1357.2--0933, Swift\,J1753.5--0127) that could be prime candidates for an origin in the Galactic halo, unless they received substantial natal kicks (typically $\gtsim$\,200--500\,km\,s$^{-1}$; see discussions in \citealt{mirabel01, gonzalezhernandez08, repetto15, belc_apj}), and these should be the focus of future dedicated studies. {\em Gaia} DR2 has already provided kinematic constraints on XTE\,J1118+480 and Swift\,J1753--0127  \citep{gandhi18}, though these constraints remain weak and should be improved upon in future releases. Another noteworthy source is VLA\,J2130+12, a likely low-mass halo BHXB candidate  ($z$\,$\gtrsim$\,1\,kpc) whose \porb\ has not been directly measured, but is likely to be short (\porb\,$\approx$\,1--2\,h), again aligning it with our observed trend \citep{tetarenko16}.

\begin{figure}
	\centering
	\hspace*{-1cm}
	\includegraphics[angle=0,width=1.1\linewidth]{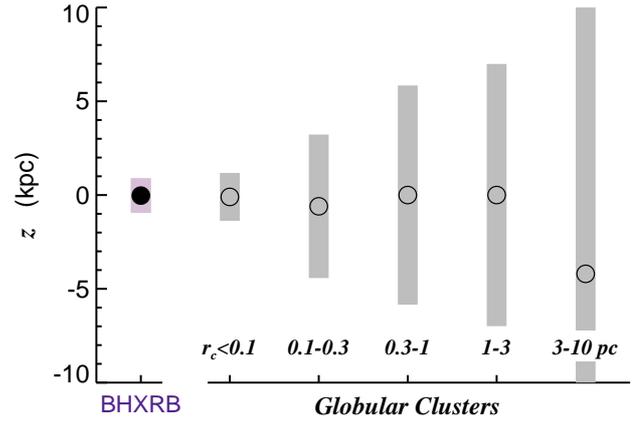}
\caption{{The circles and uncertainties denote the median $z$ and the scatter, respectively, for detected BHXBs compared against those of GCs. The GC sample is split into 0.5\,dex rangesaccording to core radius ($r_c$), as labelled. 
}}
	\label{zscatter}
	\vspace*{-0.3cm}
\end{figure}

To test a GC origin on the overall sample, we compared the BHXB \z{}-distribution with that of the GCs. Milky Way GC data were drawn from \citet[][2010 edition]{harris}. There are 155 GCs in this catalogue with measurements of core radii, spanning a vast range of distances out to 125\,kpc. Their median scatter (standard deviation) in elevation is $z^{\rm GC}_{\rm rms}$\,=\,14.5\,kpc. By contrast, we have a much smaller  $z^{\rm BHXB}_{\rm rms}$\,=\,0.9\,kpc for the 19 BHXBs with definite measured values from Fig.\,1. GCs tend to undergo tidal evaporation of periphery stars as they migrate towards the Galactic centre at disc crossings. This results in smaller elevations and more compact core radii (\rcore) for GCs deeper within the Galactic potential. It is thus natural to ask whether GCs split according to \rcore\ could better correspond to BHXBs. The result is shown in Fig.\,\ref{zscatter}, where the GC catalogue is divided into 0.5\,dex unit subgroups in \rcore\ between 0.1--10 pc. As expected, $z^{\rm GC}_{\rm rms}$ increases (from 1.3\,kpc to 28.8\,kpc) with increasing \rcore\ over the subgroups. The scatter in $z$ for the most compact GC subgroup (with \rcore\,$\lesssim$\,0.1\,pc) agrees much better with BHXBs.  The most compact clusters are the ones most likely to contain X-ray sources \citep{bellazzini95}, but these are not necessarily the clusters most likely to contain black holes \citep{ye2019}.

Two clear observational tests for GC origin are plausible with additional data.  First, the abundances of the donor stars in the short \porb\ systems should provide evidence of their origins. Only a small sample of these objects have currently been studied, and there are no indications of subsolar abundances in any of the XRB donors  \citep{2017hsn..book.1499C}.  Some $\alpha$-element enhancements have been seen and these are often argued to be evidence for supernova pollution. More detailed analysis of these abundances is thus warranted. Second, if a substantial fraction of the large scale-height systems are produced in clusters, a substantial fraction should have retrograde orbits.  At the present time, not enough X-ray binaries have good proper motion measurements to test this idea, but with the Next Generation Very Large Array, such tests would be possible \citep{2018ASPC..517..711M}.  For natal kicks to produce X-ray binaries with retrograde motion would require kicks to the binary of at least 220 km/sec which would have to be finely tuned in direction, or much faster kicks if not finely tuned in direction.

% ******************************************************

\subsubsection{The absence of outbursting BHXBs in Milky Way GCs}
Further modelling would be needed to test such a scenario quantitatively. But we point out that any successful model must additionally explain a curious fact regarding Milky Way GCs: It has been known from the early days of X-ray astronomy that GCs can host bright transient XRBs. In fact, more than 15 confirmed bright NSXRBs are known in Galactic GCs \citep{arash14}. By contrast, there is not a single confirmed case of an outbursting BHXB in a Galactic GC. There are now a few candidate quiescent systems (cited in the Introduction) 
but given the statistics of bright transient NSXRBs, one would have expected many more, and brighter, transient BHXBs. Specifically, the ratio of BH--to--NS confirmed and candidate field transients is $\approx$\,1:1 in the XRB catalogue of \citet{liu07}, so one would naively expect $\sim$\,15 outbursting BHXBs in Milky Way GCs, all else being equal, whereas none are known.

This discrepancy cannot be explained if BHs and NSs receive kicks of similar magnitude \citep[e.g. ][]{repetto15}, as this would lead to similar ejection efficiencies for both classes from GCs. The presence/absence of BHXBs in GCs has been debated on various grounds e.g. mass segregation process leading to BH decoupling \citep{lyman69} or ejection of nearly all BHs on a relatively short time-scale caused by rapid dynamical evolution in clusters of high central density \citep{kulkarni93}. It is plausible that observational biases are at play, such as lack of spatial resolution or outburst detectability, at large distances ($\sim$\,tens of kpc). However, neither of these has prevented the identification of transient GC NSXBs. \newline

\noindent
As a final note, our discussion is not meant to favour solely one or the other scenario. A combination is not ruled out, neither is an origin in the Bulge, where studies are ongoing \citep{shaw20_bulge}. 

\vspace*{-0.2cm}
\noindent
\section*{Acknowledgments}
We gratefully acknowledge prompt comments by the reviewer, and suggestions by A. Bahramian. PG acknowledges support from STFC (ST/R000506/1) and a UGC-UKIERI Thematic Partnership, and comments from C.\, Heinke and P.\,Atri. AR was in receipt of a CSC Fellowship. KB acknowledges support from the Polish National Science Center (NCN) grant. The \gaia\ mission is funded by the European Space Agency.

\vspace*{-0.5cm}
\bibliographystyle{mnras}
\bibliography{main} % if your bibtex file is called example.bib

% Don't change these lines
%\bsp	% typesetting comment
%\label{lastpage}
\label{lastpage}

%******************************************************
\clearpage
\section{Online Appendix}
Table\,\ref{table:tab2} lists the orbital period (\porb) and the Galactic height (kpc), together with references. The sample primarily comes from BlackCAT \citep{blackcat}, with a few additional comparison sources listed. Updated references are listed in all cases, where relevant.
\newpage
{%\tiny
\begin{table*}
\begin{tabular}{|>{\tiny}l>{\tiny}c>{\tiny}c>{\tiny}c>{\tiny}r|}
\hline
\hline
Name & \porb & $z$ & Reference (\porb) & Reference (Distance)\\
     & (hour) & (kpc)\\
\hline
4U\,1543--475 & 26.79377\,(7) & 0.71\,$\pm$\,0.05 & \citet{orosz03} & \citet{jonker04}\\
4U\,1755--338 & 4.4\,(1.0) & --0.55\,$\pm$\,0.21 & \citet{white84} & \citet{angelini03}\\
GX\,339--4 & 42.14\,(1) & <\,--0.38 & \citet{hynes03} & \citet{heida}\\
3A\,0620--003 & 7.7523372(2) & -0.12\,$\pm$\,0.01 &\citet{gonz10} & \citet{cantrell10}\\
H\,1705--250 & 12.51(3) & 1.35\,$\pm$\,0.33 &\citet{remi96} & \citet{jonker04}\\
BW\,Cir & 61.068(5) & <\,--1.21 & \citet{cas09} & \citet{gandhi18}\\
GS\,2000+251 & 8.25821(2) & --0.14\,$\pm$\,0.04 &\citet{cas95} & \citet{jonker04}\\
V404\,Cyg & 155.30808(5) & --0.09\,$\pm$\,0.01 &\citet{casares19} & \citet{millerjones09}\\
GRS\,1124--684 & 10.38254(7) & --0.73\,$\pm$\,0.03 &\citet{orosz96} & \citet{hynes05}\\
GRO\,J0422+32 & 5.091840(5) & --0.51\,$\pm$\,0.06 &\citet{webb} & \citet{gelino}\\
GRS\,1915+105 & 812(4) & --0.03\,$\pm$\,0.01 &\citet{steeghs} & \citet{reid14}\\
GRS\,1009--45 & 6.84494(3) & 0.62\,$\pm$\,0.05 &\citet{fili99} & \citet{gelino02}\\
GRO\,J1655--40 & 62.920(3) & 0.14\,$\pm$\,0.01 &\citet{van98} & \citet{hjell95}\\
XTE\,J1550--564 & 37.00880(6) & --0.14\,$\pm$\,0.02 &\citet{orosz11} & \citet{orosz11}\\
SAX\,J1819.3--2525 & 67.6152(2) & --0.52\,$\pm$\,0.06 &\citet{orosz01} & \citet{mac14}\\
XTE\,J1859+226 & 6.58(5) & 0.63\,$\pm$\,0.07 &\citet{corral11} & \citet{shapo09}\\
XTE\,J1118+480 & 4.078414(5) & 1.52\,$\pm$\,0.09 &\citet{torres04} & \citet{gelino06}\\
XTE\,J1650--500 & 7.69(2) & --0.16\,$\pm$\,0.04 &\citet{orosz04} & \citet{homan06}\\
SWIFT\,J1753.5--0127 & 3.244(1) & 1.27\,$\pm$\,0.42 & \citet{zurita08} & \citet{cado07}\\
XTE\,J1752--223 & <6.8 & 0.22\er{}0.07 & \citet{ratti12} & \citet{ratti12}\\
MAXI\,J1659--152 & 2.414(5) & 2.4\,$\pm$\,1.0 & \citet{kuulkers} & \citet{kuulkers}\\
SWIFT\,J1357.2--0933 & 2.8(3) & >4.60 &\citet{corral2013} & \citet{charles19}\\
MAXI\,J1836--194 & <4.9 & --0.65\,$\pm$\,0.28 & \citet{russell14} & \citet{russell14}\\
SWIFT\,J174510.8--262411 & $\lesssim$21 & <0.17 & \citet{munoz2013} & \citet{munoz2013}\\
MAXI\,J1820+070 & 16.45(2) & 0.52\,$\pm$\,0.06 & \citet{torres19} & \citet{atri20}\\
&&&&\\
SS\,433 & 313.974(1) & --0.215(8) &\citet{goranskij11} & \citet{blundell04}\\
Cyg\,X--1 & 134.395(3) & 0.13\,$\pm$\,0.01 &\citet{lasala98} & \citet{rao19, gandhi18};\\
& && & \citet{reid11,mirabelrodrigues03}\\
Cyg\,X--3 & 4.793(1) & 0.09$_{-0.02}^{+0.03}$ &\citet{singh02} & \citet{mccollough16}\\
VLA\,J213002.08+120904 & 1.5(5) & --1.01$_{-0.09}^{+0.14}$ &\citet{tetarenko16} & \citet{kirsten14}\\
\hline
\end{tabular}
\caption{Orbital periods and \z{}-heights. Some of the primary references are listed, with additional work cited in BlackCAT. The brackets in column 2 denote uncertainties in the last digits. The final four systems are used as comparison, and were not included in our statistical fits. 
}
\label{table:tab2}
\end{table*}
}
  %******************************************************

\end{document}